\documentclass[12pt]{iopart}

\usepackage{graphicx}
\usepackage{color}
\begin{document}

\title{Exact Solutions and Quantum Defect Theory for van der Waals Potentials in Ultracold Molecular Systems}

\author{Jianwen Jie$^{1}$$^{*}$, Shi Chen$^{1}$$^{*}$, Yue Chen$^{2,3}$$^{\dagger}$ and Ran Qi$^{4}$$^{\dagger}$}
\address{$^1$Shenzhen Key Laboratory of Ultraintense Laser and Advanced Material Technology, Center for Intense Laser Application Technology, and College of Engineering Physics, Shenzhen Technology University, Shenzhen, 518118, China}
\address{$^2$ China Electronics Corporation, Shenzhen, 518057, China}
\address{$^3$ China Great wall Technology Group Co., Ltd, Shenzhen 518000, China}
\address{$^4$ Department of Physics, Renmin University of China, Beijing, 100872, China}
\eads{\mailto{yuechenphy@163.com}, \mailto{qiran@ruc.edu.cn}}

\vspace{10pt}
\begin{indented}
\item[] $^*$ These authors contributed equally to this work.
\item[] $^\dagger$ Author to whom any correspondence should be addressed.
\end{indented}

\begin{abstract}
In this paper, we have provided exact two-body solutions to the 2D and 3D Schrödinger equations with isotropic van der Waals potentials of the form \(\pm 1/r^6\). Based on these solutions, we developed an analytical quantum defect theory (QDT) applicable to both quasi-2D and 3D geometries, and applied it to study the scattering properties and bound-state spectra of ultracold polar molecules confined in these geometries. Interestingly, we find that in the attractive (repulsive) van der Waals potential case, the short-range interaction can be effectively modeled by an infinite square barrier (finite square well), which leads to narrow and dense (broad and sparse) resonance structures in the quantum defect parameter. In the quasi-2D attractive case, shape resonances can appear in an ordered fashion across different partial waves, characterized by sharp phase jumps as the scattering energy is varied. Furthermore, the low-energy analytical expansions derived from QDT show excellent agreement with the exact numerical results, validating the accuracy and usefulness of our analytical approach in describing two-body physics governed by long-range van der Waals interactions.
\end{abstract}
\maketitle

\section{Introduction}
 Systems with long-range interactions serve as essential platforms for exploring non-local correlations and novel many-body quantum phases \cite{defenu2021longrange}. Platforms for realizing diverse long-range interactions range from trapped ions \cite{Schneider_2012}, optical cavities \cite{optical_cavity}, and Rydberg atoms \cite{schauss2012observation} to polar molecular gases \cite{ni2008high}. In particular, polar molecular gases are excellent candidates for advancing the field of quantum chemistry due to their controllability and experimental accessibility \cite{jin2012introduction,liu2021bimolecular,Carroll2025Science, Langen2025PRL, Stevenson2024PRL, He2024PRL, Karman2024NatPhys, Cornish2024NatPhys, Langen2024NatPhys}.

However, two molecules can collide and overcome potential barriers, initiating complex chemical reactions that significantly shorten the lifetime of the molecular gases \cite{ospelkaus2010quantum,ye2018collisions,hu2019direct}. After decades of effort, long-lived ultracold polar KRb molecular gases exhibiting strong electric dipole-dipole interactions have recently been realized by confining the gases to a quasi-two-dimensional (2D) geometry and employing electric fields to suppress the two-body loss rate \cite{matsuda2020resonant}. Interestingly, by tuning the electric fields, both repulsive and attractive interactions between two molecules can emerge \cite{matsuda2020resonant}. Depending on the strength of the applied field and the inter-molecular separation, the interaction may exhibit either a dipolar character ($1/r^3$) or a van der Waals character ($1/r^6$). These results were later extended to fully three-dimensional (3D) settings in subsequent experiments \cite{li2021tuning}. Despite these experimental advancements, analytical results for van der Waals interactions remain limited. In two dimensions, exact solutions for both attractive and repulsive potentials are lacking. In three dimensions, while the attractive case has been solved analytically \cite{Gao1998C6}, the repulsive case remains largely unexplored.

Motivated by these developments, we present exact solutions to the Schrödinger equation for both attractive and repulsive $1/r^6$ potentials in 2D, as well as for the repulsive $1/r^6$ potential in 3D. These solutions are obtained using the generalized Neumann expansion method, previously developed for scattering problems involving dipolar interactions \cite{Jie_2016,Bo1999PRAC3} and attractive van der Waals interactions in 3D \cite{Gao1998C6}. Based on these solutions, we construct an analytic quantum defect theory (QDT) \cite{QDT_PRA_1979,QDT_PRA_1982,QDT_PRA_1984,QDT_PRA_2005,QDT_PRA_2008,QDT_PRA_2009,Gao1998C6QDT} for ultracold polar molecular gases \cite{matsuda2020resonant,li2021tuning} under quasi-2D confinement and in 3D geometry. We further analyze the resulting scattering properties and two-body bound states, providing a unified theoretical framework that complements and supports recent experimental findings.

In Sec. \ref{solutions}, we formulate the problem and present the exact solutions of the Schrödinger equation. We also provide analytical asymptotic expressions in both the short-range and long-range limits. In Sec. \ref{QDTall}, we first examine the two-body problem in quasi-2D and 3D confined geometries. Based on this, we develop analytical quantum defect theories (QDTs) for van der Waals interactions, and apply them to investigate the corresponding scattering properties and bound states. We conclude the paper in Sec. \ref{conclusion}.

\section{Solutions of the Schr\"{o}dinger equation}\label{solutions}
We analyze the radial Schrödinger equation with van der Waals potentials of the form $\pm C_6 / r^6$ in two and three dimensions, governing the radial wave function $\overline{u}_{\overline{\epsilon} \overline{l}}(r)$ {\color{black}(see the \ref{ascasd} for the detailed derivation)}:
\begin{equation}
\left[ \frac{d^{2}}{d r^{2}} - \frac{\overline{l}(\overline{l}+1)}{r^{2}} - \delta\frac{\beta_{6}^{4}}{r^6} + \overline{\epsilon} \right] \overline{u}_{\overline{\epsilon} \overline{l}}(r) = 0,
\label{Schrodinger}
\end{equation}
where $\overline{\epsilon}$ and $\overline{l}$ are the reduced energy and effective angular momentum, respectively.  Here, $\beta_6 = (2\mu C_6 / \hbar^2)^{1/4}$ defines the van der Waals length, with $\mu$ and $C_6$ being the reduced mass and interaction strength, respectively. The reduced energy is defined as $\overline{\epsilon} = 2\mu \epsilon / \hbar^2 = k^2$, where $\epsilon$ is the scattering energy and $k$ is the relative wave number.  In 2D, $\overline{l} = m - 1/2$, where $m$ is the azimuthal quantum number; in 3D, $\overline{l} = l$, where $l$ is the orbital angular momentum quantum number. The index $\delta$ distinguishes between repulsive ($\delta = +1$) and attractive ($\delta = -1$) potentials.

\begin{table}[b]
    \centering
    \resizebox{0.8\textwidth}{!}{%
    \small
    \begin{tabular}{|c|c|c|c|c|}
    \hline
                      & \multicolumn{2}{c|}{Attractive: $-C_{6}/r^6$} & \multicolumn{2}{c|}{Repulsive: $+C_{6}/r^6$} \\ \cline{2-5}
                      & 2D                         & 3D~\cite{Gao1998C6}   & 2D                         & 3D                         \\ \hline
    Angular momentum $\overline{l}$ & $m - \frac{1}{2}$            & $l$               & $m - \frac{1}{2}$              & $l$               \\ \hline
    $L$               & $\beta_{6}\sqrt{\frac{1}{2}}$ & $\beta_{6}\sqrt{\frac{1}{2}}$ & $\beta_{6}\sqrt{\frac{i}{2}}$ & $\beta_{6}\sqrt{\frac{i}{2}}$ \\ \hline
    \end{tabular}%
    }
    \caption{Summary of the definitions of the effective angular momentum quantum number $\overline{l}$ and the length scale $L$ for all attractive and repulsive van der Waals potentials in 2D and 3D geometries.}
    \label{table1}
\end{table}

{\color{black}In this work, we follow the approach first developed by Gao in his exact solutions to the 3D attractive van der Waals potential and the repulsive dipole-dipole interaction~\cite{Gao1998C6,Bo1999PRAC3}. To derive analytical solutions, we begin by transforming Eq.~(\ref{Schrodinger}) into a dimensionless form using the following substitutions:}
\begin{equation}\label{ru}
r = \frac{L}{\sqrt{x}}, \quad \overline{u}_{\overline{\epsilon}\overline{l}}(r) = \sqrt{r}\,f(x),
\end{equation}
where $x$ and $f(x)$ denote a dimensionless coordinate and wave function, respectively, and $L = \beta_{6} (\sqrt{\delta}/2)^{1/2}$. This yields the following differential equation:
\begin{equation}\label{xfde}
x^{2}f''(x) + x f'(x) + (x^{2} - \nu_{0}^{2}) f(x) = -\frac{2\Delta}{x} f(x), \label{Schrodinger1}
\end{equation}
where $\Delta = \overline{\epsilon} L^2 / 8$ is the scaled energy and $\nu_0 = (2\overline{l} + 1)/4$. For the attractive van der Waals potential in three dimensions, the solution to Eq.~(\ref{Schrodinger1}) was obtained in Ref.~\cite{Gao1998C6} using the Neumann expansion method~\cite{Ca1994PRA_NE,abramowitz1964handbook,watson1995treatise}. This method has also been extended to dipole interactions ($\pm C_3/r^3$) in both 2D~\cite{Jie_2016}. This approach allows us to derive solutions in a unified manner for both attractive and repulsive van der Waals potentials in 2D and 3D, with the corresponding definitions of the length scale $L$ and the effective angular momentum quantum number summarized in Table~\ref{table1}. {\color{black}It is worth noting that our Eq.~(\ref{ru}) and Eq.~(\ref{xfde}) are formally identical to Eqs.~(32-34) in Gao's treatment of the 3D attractive van der Waals potential~\cite{Gao1998C6}. However, the key difference lies in the definition of the length scale \( L \), as summarized in Table~\ref{table1}, and the associated scaled energy parameter \( \Delta \). These distinctions lead to a different set of exact solutions and corresponding asymptotic behaviors.} We now present the explicit forms of the exact solutions to Eq.~(\ref{Schrodinger}) in the following part of this section. {\color{black}The results obtained in Secs.~2.1 and 2.2 are valid for all cases listed in Table~\ref{table1}.}

\subsection{Summary of the generalized Neumann expansion solutions}

{\color{black}By solving the second-order differential equation Eq.~(\ref{xfde}) and substituting the results into Eq.~(\ref{ru}), we obtain a pair of linearly independent solutions of Eq. (\ref{Schrodinger}), denoted as \( \overline{u}_{\overline{\epsilon} \, \overline{\ell}}^{(1)}(r) \) and \( \overline{u}_{\overline{\epsilon} \, \overline{\ell}}^{(2)}(r) \),} which can be expressed in the form of generalized Neumann expansions:
\begin{eqnarray}
\overline{u}_{\overline{\epsilon}\overline{l}}^{1}(r) &=& r^{1/2}\sum_{n=-\infty}^{\infty}b_{n}J_{\nu+n}\left(\frac{L^{2}}{r^{2}}\right),\label{ubar1}\\
\overline{u}_{\overline{\epsilon}\overline{l}}^{2}(r) &=& r^{1/2}\sum_{n=-\infty}^{\infty}b_{n}Y_{\nu+n}\left(\frac{L^{2}}{r^{2}}\right),\label{ubar2}
\end{eqnarray}
where the coefficients are given by
\begin{eqnarray}
&&b_{j}=(-\Delta)^{j}\frac{\Gamma(\nu)\Gamma(\nu-\nu_{0}+1)\Gamma(\nu+\nu_{0}+1)}{\Gamma(\nu\!+\!j)\Gamma(\nu-\nu_{0}+j+1)\Gamma(\nu+\nu_{0}+j+1)}c_{j}(\nu),~~\\
&&b_{-j}=(-\Delta)^{j}\frac{\Gamma(\nu-j+1)\Gamma(\nu-\nu_{0}-j)\Gamma(\nu+\nu_{0}-j)}{\Gamma(\nu\!+1)\Gamma(\nu-\nu_{0})\Gamma(\nu+\nu_{0})}c_{j}(-\nu),~~
\end{eqnarray}
with \(j\) being a positive integer, $\nu_0=(2\bar{l}+1)/4$ and the function \(c_j\) defined as
\begin{equation}
c_{j}(\nu) = Q(\nu+j-1)Q(\nu+j-2)\cdots Q(\nu) b_0.
\end{equation}
Here, \(b_0\) is a normalization constant (which can be set to 1), and \(Q(\nu)\) is given by a continued fraction:
\begin{equation}\label{expereQ}
Q(\nu) = \frac{1}{1 - \Delta^2 \frac{Q(\nu+1)}{(\nu+1)[(\nu+1)^2 - \nu_0^2](\nu+2)[(\nu+2)^2 - \nu_0^2]}}.
\end{equation}
The index parameter \(\nu\) appearing in Eqs.~(\ref{ubar1}--\ref{expereQ}) is a root of the characteristic function
\begin{equation}
\Lambda_{\overline{l}}(\nu, \Delta^2) \equiv \left(\nu^2 - \nu_0^2\right) - \frac{\Delta^2}{\nu} \left[\overline{Q}(\nu) - \overline{Q}(-\nu)\right],\label{Lambdaeq}
\end{equation}
where
\begin{equation}
\overline{Q}(\nu) = \frac{Q(\nu)}{(\nu+1)[(\nu+1)^2 - \nu_0^2]}.
\end{equation}
The solution \(\nu\) to the equation \(\Lambda_{\overline{l}}(\nu, \Delta^2) = 0\) can be either real or complex, depending on the scattering energy and angular momentum. The determination of \(\nu\) is a key step in constructing the exact solution. A detailed analysis of the energy dependence of \(\nu\), as well as its low-energy expansion for different partial waves, will be presented in Sec.~\ref{sec_nu}.

{\color{black}We find that there exists a pair of linearly independent and real solutions with energy-independent asymptotic behavior near the origin ($r \ll \beta_6$). These solutions for the replusive potentials can be expressed explicitly as}
\begin{eqnarray}
u_{\overline{\epsilon}\overline{l}}^{+1}(r) &\!=\!& \widetilde{X}^{-1}\left[i \overline{u}_{\overline{\epsilon}\overline{l}}^{1}(r)-\overline{u}_{\overline{\epsilon}\overline{l}}^{2}(r)\right],\label{up1}\\
u_{\overline{\epsilon}\overline{l}}^{+2}(r) &\!=\!& \widetilde{Y}^{-1}\overline{u}_{\overline{\epsilon}\overline{l}}^{1}(r),\label{up2}
\end{eqnarray}
{\color{black}and for the attactive potentials, we have}
\begin{eqnarray}
u_{\overline{\epsilon}\overline{l}}^{-1}(r) &\!=\!& \left(\alpha^{2}+\beta^{2}\right)^{-1}\left[\alpha \overline{u}_{\overline{\epsilon}\overline{l}}^{1}(r)-\beta  \overline{u}_{\overline{\epsilon}\overline{l}}^{2}(r)\right],\label{un1}\\
u_{\overline{\epsilon}\overline{l}}^{-2}(r) &\!=\!& \left(\alpha^{2}+\beta^{2}\right)^{-1}\left[\beta \overline{u}_{\overline{\epsilon}\overline{l}}^{1}(r)+\alpha \overline{u}_{\overline{\epsilon}\overline{l}}^{2}(r)\right].\label{un2}
\end{eqnarray}
The normalization coefficients in Eqs.~(\ref{up1}--\ref{un2}) are given by
\begin{eqnarray}
\left(
\begin{array}{c}
\alpha \\
\beta 
\end{array}
\right)
= 
\left(
\begin{array}{cc}
-Y & X \\
X & Y
\end{array}
\right)
\left(
\begin{array}{c}
\sin\left[\frac{\pi}{2}(\nu - \nu_0)\right] \\
\cos\left[\frac{\pi}{2}(\nu - \nu_0)\right]
\end{array}
\right), \label{ab}
\end{eqnarray}
and
\begin{eqnarray}
\widetilde{X} = \sqrt{2}i^{-\nu}(X - iY), \quad \widetilde{Y} = \frac{i^{\nu}(X + iY)}{\sqrt{2}},\label{xy}
\end{eqnarray}
where
\begin{equation}
X = \sum_{n=-\infty}^{\infty}(-1)^n b_{2n}, \quad Y = \sum_{n=-\infty}^{\infty}(-1)^n b_{2n+1}.
\end{equation}

\subsection{Asymptotic behavior}
The solution pairs $u_{\overline{\epsilon}\overline{l}}^{\pm1}(r)$ and $u_{\overline{\epsilon}\overline{l}}^{\pm2}(r)$ are constructed such that they exhibit energy-independent behavior in the short-range limit ($r \ll \beta_6$). Their asymptotic forms near the origin are given by
\begin{eqnarray}
u_{\overline{\epsilon}\overline{l}}^{+1}(r) &\rightarrow& \frac{r}{\beta_6} \sqrt{\frac{2r}{\pi}}\, e^{-\beta_6^2/2r^2},\label{u1pasym}\\
u_{\overline{\epsilon}\overline{l}}^{+2}(r) &\rightarrow& \frac{r}{\beta_6} \sqrt{\frac{2r}{\pi}}\, e^{\beta_6^2/2r^2},\label{u2pasym}\\
u_{\overline{\epsilon}\overline{l}}^{-1}(r) &\rightarrow& \sqrt{\frac{2r}{\pi}}\, \frac{r}{L} \cos\left(\frac{L^2}{r^2} - \frac{\nu_0\pi}{2} - \frac{\pi}{4} \right),\label{u1nasym}\\
u_{\overline{\epsilon}\overline{l}}^{-2}(r) &\rightarrow& \sqrt{\frac{2r}{\pi}}\, \frac{r}{L} \sin\left(\frac{L^2}{r^2} - \frac{\nu_0\pi}{2} - \frac{\pi}{4} \right),\label{u2nasym}
\end{eqnarray}
and are valid for both positive and negative energies. For positive energy $\overline{\epsilon} = k^2 > 0$, the asymptotic behavior of the solutions as $r \rightarrow +\infty$ is
\begin{equation}\label{u1asymlongp}
\left(
\begin{array}{c}
u_{\overline{\epsilon}\widetilde{l}}^{\pm1}(r)  \\
u_{\overline{\epsilon}\overline{l}}^{\pm2}(r)
\end{array}
\right)
\rightarrow  \sqrt{\frac{2}{\pi k}}
\left(
\begin{array}{cc}
Z_{11}^{\pm} & -Z_{12}^{\pm} \\
Z_{21}^{\pm} & -Z_{22}^{\pm}
\end{array}
\right)
\left(
\begin{array}{c}
\sin(kr - \widetilde{l}\pi/2) \\
\cos(kr - \widetilde{l}\pi/2)
\end{array}
\right),
\end{equation}
{\color{black}where the index \(\widetilde{l}\) can in principle be any integer. However, to maintain consistency with the standard asymptotic form of the scattering wavefunction at large distances, we set \(\widetilde{l} = l\) in 3D and \(\widetilde{l} = m\) in 2D.} For negative energy $\overline{\epsilon} = k^2 = -\kappa^2 < 0$, the large-$r$ asymptotic behavior becomes
\begin{equation}\label{u1asymlongn}
\left(
\begin{array}{c}
u_{\overline{\epsilon}\widetilde{l}}^{\pm1}(r)  \\
u_{\overline{\epsilon}\overline{l}}^{\pm2}(r)
\end{array}
\right)
\rightarrow  \sqrt{\frac{2}{\pi \kappa}}
\left(
\begin{array}{cc}
W_{11}^{\pm} & W_{12}^{\pm} \\
W_{21}^{\pm} & W_{22}^{\pm}
\end{array}
\right)
\left(
\begin{array}{c}
e^{\kappa r} \\
e^{-\kappa r}
\end{array}
\right).
\end{equation}
From the small-$r$ asymptotic forms in Eqs.~(\ref{u1pasym})--(\ref{u2nasym}), it is straightforward to verify that the Wronskian of each solution pair satisfies
\begin{equation}
W(u_{\overline{\epsilon}\overline{l}}^{\pm1}, u_{\overline{\epsilon}\overline{l}}^{\pm2}) = -\frac{4}{\pi}.
\end{equation}
Since the Wronskian is independent of $r$, its value must also be preserved in the large-$r$ limit. This imposes the following constraints on the transformation matrices:
\begin{eqnarray}
\det(Z^{\pm}) &=& Z_{11}^{\pm} Z_{22}^{\pm} - Z_{12}^{\pm} Z_{21}^{\pm} = -2, \\
\det(W^{\pm}) &=& W_{11}^{\pm} W_{22}^{\pm} - W_{12}^{\pm} W_{21}^{\pm} = 4.
\end{eqnarray}
These relations are independent of the energy $\overline{\epsilon}$ and angular momentum $\overline{l}$, and hold universally for all cases listed in Table~\ref{table1}. {\color{black}Physically, this reflects the conservation of probability current in quantum mechanics, as the Wronskian is proportional to the flux carried by the wavefunctions. In scattering problems, this invariance ensures that boundary-matching conditions preserve the normalization and phase coherence of asymptotic solutions.} We have verified them in our calculations as a nontrivial consistency check of the solution structure. The explicit results for the $W^{\pm}$ and $Z^{\pm}$ matrices are provided in \ref{WZ}.

\section{Quantum defect theory of the van der Waals potentials in quasi-2D and 3D}\label{QDTall}

One usual example to understanding the quantum defect theory is the energy spectrum of hydrogen atom and alkali atoms \cite{seaton1983quantum}. For hydrogen atom, there is only one electron moving around the nucleus, and the energy spectrum is simply proportional to $1/n^{2}$, with $n$ being the principle quantum number. While for the alkali atoms, in addition to the outermost electron, there are still inner-shell electrons which will provide the penetration effect and screening effect when the outermost electron moving around the nucleus. Therefore, one need to introduce a parameter $\Delta$ for the energy spectrum of alkali atoms to describe those short range effects and then the modified energy spectrum could be simply proportional to $1/(n-\Delta)^{2}$ compared to the one of the hydrogen atom. The introduced parameter $\Delta$ contains the information of the short range potential of the outermost electron and usually be called as quantum defect parameter.  

For ultracold and dilute molecular gases, the average distance between two { molecules} is much larger than the {van} der Waals length and then we usually can model the two-body interaction by the contact interaction associated with the scattering length which contains the information of short range potentials \cite{RMP_FR}. Either when one wants to have more accurate two-body scattering properties or when the molecular gases are not enough cold and dilute, we need to consider the { van} der Waals type shorter range potential, which can be induced from two polar molecules tuning by electric fields \cite{matsuda2020resonant,li2021tuning}. {In these experiments \cite{matsuda2020resonant,li2021tuning},} two $^{40}$K$^{87}$Rb molecules prepared in the {\color{black} the rovibrational state $|\nu=0,N=1,m_{N}=0 \rangle=|1,0\rangle$ of the singlet electronic potential $X^{1}\Sigma$ \cite{Ni2008}, where $\nu$ is the vibrational quantum number}, $N$ is the rotational angular momentum and $m_{N}$ is its projection along the axis of electric field ${\bf E}$.  Starting from this two-molecular state $|1,0\rangle\otimes|1,0\rangle$ at weak  electric field ${\bf E}$, {  they increased} the strength of electric field $|{\bf E}|$ to across two crossing energy points, where the two molecular states are  $|0,0\rangle\otimes|2,\pm1\rangle$  and  $|0,0\rangle\otimes|2,0\rangle$, respectively.  Near the crossing points, the two molecular states are coupled by the dipole interaction which weakly opens the energy gap via a second order perturbation process. Thus the effective attractive (repulsive) { van} der Waals barrier in the energy curve of these two molecules associated with state $|0,0\rangle\otimes|2,\pm1\rangle$ ($|0,0\rangle\otimes|2,0\rangle$) emerges.  This new technique for inducing the effective { van} der Waals potentials both be realized in quasi-2D  \cite{matsuda2020resonant}  and 3D  \cite{li2021tuning}. Next, we will construct the quantum defect theories for these systems.
\begin{figure}[t]
\centering
\includegraphics[width=12cm]{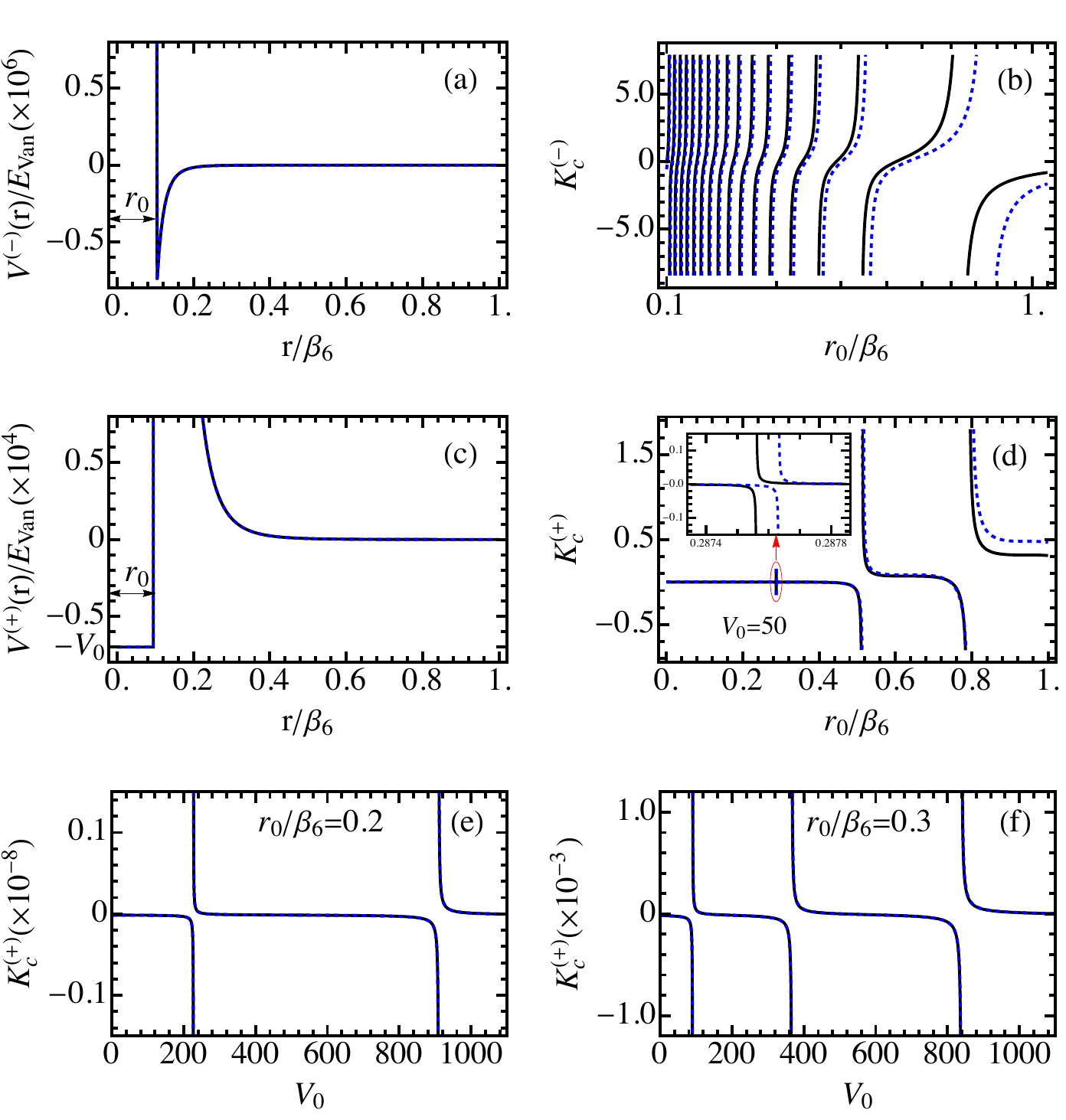}   
\caption{(Color online): Quantum defect parameter for two potential configurations $V^{(-)}$ (a-b) and  $V^{(+)}$ (c-f). The black solid lines are for 3D and the dashed blue lines are for 2D. In (a-b), all cases are $s$ wave. In (c-f),  $s$ wave and $p$ wave are picked up for 3D and 2D cases, respectively.
}
\label{Kcplot}
\end{figure}

\subsection{Quantum defect parameter $K_c$}
In Eqs.~(\ref{up1})--(\ref{un2}), we introduced a pair of linearly independent specific solutions. The general solution can be expressed as a linear combination of these solutions:
\begin{eqnarray}
u_{\overline{\epsilon}\overline{l}}^{\pm}(r) = A \left[ u_{\overline{\epsilon}\overline{l}}^{\pm1}(r) - K_c^{(\pm)} u_{\overline{\epsilon}\overline{l}}^{\pm2}(r) \right],
\end{eqnarray}
where the relative amplitude \(K_c^{(\pm)}\) is referred to as the \emph{quantum defect parameter}, which encodes the short-range information of the system.

At positive scattering energies, the phase shift can be extracted by comparing the asymptotic forms of \(u_{\overline{\epsilon}\overline{l}}^{\pm1}(r)\) and \(u_{\overline{\epsilon}\overline{l}}^{\pm2}(r)\) given in Eq.~(\ref{u1asymlongp}):
\begin{eqnarray}
\mbox{3D:} \quad \tan \delta_l^{(\pm)} &=& \frac{K_c^{(\pm)} Z_{22}^{(\pm)} - Z_{12}^{(\pm)}}{Z_{11}^{(\pm)} - K_c^{(\pm)} Z_{21}^{(\pm)}} \label{ps_de1} \\
\mbox{2D:} \quad \tan \delta_m^{(\pm)} &=& \frac{K_c^{(\pm)} (Z_{21}^{(\pm)} + Z_{22}^{(\pm)}) - (Z_{11}^{(\pm)} + Z_{12}^{(\pm)})}{(Z_{11}^{(\pm)} - Z_{12}^{(\pm)}) - K_c^{(\pm)} (Z_{21}^{(\pm)} - Z_{22}^{(\pm)})} \label{ps_de2}
\end{eqnarray}
The differential scattering cross sections are related to the phase shifts as follows:
\begin{eqnarray}
\mbox{3D:} \quad \sigma_l^{(\pm)} &=& 4\pi(2l+1)\frac{\sin^2 \delta_l^{(\pm)}}{k^2}, \label{sec_de1} \\
\mbox{2D:} \quad \sigma_m^{(\pm)} &=& 4 \frac{\sin^2 \delta_m^{(\pm)}}{k}. \label{sec_de2}
\end{eqnarray}
For bound states, the physical requirement is that the wave function decays exponentially at large \(r\). This means the coefficient of the exponentially growing term \(e^{\kappa r}\) in the asymptotic expansion of \(u_{\overline{\epsilon}\overline{l}}^{\pm}(r)\) [see Eq.~(\ref{u1asymlongn})] must vanish. This yields the bound-state condition:
\begin{equation}
W_{11}^{(\pm)} - K_c^{(\pm)} W_{21}^{(\pm)} = 0.
\end{equation}
Accordingly, we define a function \(\chi^{(\pm)}_{\widetilde{l}}(E_b)\) to characterize the bound-state condition:
\begin{equation}
\chi^{(\pm)}_{\widetilde{l}}(E_b) = K_c^{(\pm)} = \frac{W_{11}^{(\pm)}}{W_{21}^{(\pm)}}. \label{chi_de}
\end{equation}

Therefore, whether one aims to determine phase shifts via Eqs.~(\ref{ps_de1})--(\ref{ps_de2}) or extract binding energies from Eq.~(\ref{chi_de}), the quantum defect parameter \(K_c^{(\pm)}\) must be specified. {\color{black}Moreover, in the short-range region \( r \ll \beta_6 \), the total energy \( \overline{\epsilon} \) of the particle is typically much smaller than the local potential energy. As a result, the wavefunctions in this region become effectively insensitive to both energy and angular momentum, as demonstrated in Eqs.~(\ref{u1pasym})--(\ref{u2nasym}). Since the quantum defect parameter \( K_c^{(\pm)} \) is determined by the logarithmic derivative of the short-range solutions, it also becomes approximately independent of energy and partial wave at short distances.} To illustrate this property, we divide the total potential into a short-range part and a long-range tail using a boundary located at \(r = r_0\). The long-range tail is governed by the van der Waals potential \(\pm C_6 / r^6\), while the short-range component is modeled using idealized potentials: a finite square well for the repulsive case, and an infinite square barrier for the attractive case. These model potentials are shown schematically in Eqs.~(\ref{Vp_de})--(\ref{Vn_de}) and illustrated in Fig.~\ref{Kcplot}(a,c), respectively.

\begin{eqnarray}
V^{(+)}(r) &=& -V_0\, \theta(r_0 - r)\, E_{\mbox{van}} + \theta(r - r_0)\, \frac{\hbar^2}{2\mu} \left[ \frac{\overline{l}(\overline{l}+1)}{r^2} + \frac{\beta_6^4}{r^6} \right], \label{Vp_de} \\
V^{(-)}(r) &=& +\infty\, \theta(r_0 - r)\, E_{\mbox{van}} + \theta(r - r_0)\, \frac{\hbar^2}{2\mu} \left[ \frac{\overline{l}(\overline{l}+1)}{r^2} - \frac{\beta_6^4}{r^6} \right], \label{Vn_de}
\end{eqnarray}
where \(E_{\mbox{van}} = \hbar^2 / (2\mu \beta_6^2)\) is the characteristic energy scale associated with the van der Waals length \(\beta_6\). The functions \(\theta(r)\) denote the Heaviside step function. Equation~(\ref{Vp_de}) represents a model for the repulsive van der Waals potential, composed of a square well of depth \(V_0 E_{\mbox{van}}\) for \(r < r_0\), while Eq.~(\ref{Vn_de}) models the attractive case using an infinite barrier for \(r < r_0\).

Based on these model potentials, and applying matching conditions at the boundary \(r = r_0\), we can analytically derive the zero-energy expressions for the quantum defect parameter \(K_c^{(\pm)}\). For the repulsive potential \(V^{(+)}\), the expression is
{\setlength{\mathindent}{0pt}
\begin{eqnarray}
K_c^{(+)} = i^{2\nu_0 + 1} \frac{
2r_0^3 \sqrt{V_0}\, H^{(1)}_{\nu_0} \left( \frac{L^2}{r_0^2} \right) \cos\left( \frac{r_0 \sqrt{V_0}}{\beta_6} \right) + 
\beta_6 \left[ 4L^2 H^{(1)}_{\nu_0 - 1} \left( \frac{L^2}{r_0^2} \right) - (4\nu_0 + 1) r_0^2 H^{(1)}_{\nu_0} \left( \frac{L^2}{r_0^2} \right) \right] \sin\left( \frac{r_0 \sqrt{V_0}}{\beta_6} \right)
}{4r_0^3 \sqrt{V_0}\, J_{\nu_0} \left( \frac{L^2}{r_0^2} \right) \cos\left( \frac{r_0 \sqrt{V_0}}{\beta_6} \right) - 
2\beta_6 \left[ (4\nu_0 + 1) r_0^2 J_{\nu_0} \left( \frac{L^2}{r_0^2} \right) - 4L^2 J_{\nu_0 - 1} \left( \frac{L^2}{r_0^2} \right) \right] \sin\left( \frac{r_0 \sqrt{V_0}}{\beta_6} \right)
},\nonumber\\ \label{Kp_zero}
\end{eqnarray}
}
and for the attractive potential \(V^{(-)}\), the expression simplifies to
\begin{equation}
K_c^{(-)} = \frac{J_{\nu_0} \left( \frac{L^2}{r_0^2} \right)}{Y_{\nu_0} \left( \frac{L^2}{r_0^2} \right)}. \label{Kn_zero}
\end{equation}

In Fig.~\ref{Kcplot}, we show the behavior of the quantum defect parameter \(K_c^{(\pm)}\) as a function of short-range potential parameters. The black solid lines correspond to 3D geometries, while the blue dashed lines represent 2D results. For the attractive cases [Fig.~\ref{Kcplot}(b)], the density of shape resonances increases rapidly as the short-range cutoff \(r_0\) approaches zero. In contrast, in the repulsive cases [Figs.~\ref{Kcplot}(d--f)], the resonances are significantly sparser and much narrower in width. In fact, \(K_c^{(+)}\) remains close to zero over most of the parameter space, and no prominent resonance emerges even as \(r_0 \rightarrow 0\), given a fixed potential depth \(V_0\). This behavior can be attributed to the presence of a large centrifugal and repulsive van der Waals barrier, which effectively suppresses the influence of the short-range region on the scattering wave function. Consequently, for the repulsive case, we will focus on the pure repulsive limit (\(r_0 \rightarrow 0\)), where \(K_c^{(+)} = 0\). For the attractive case, however, both scattering and bound-state properties will be analyzed across shape resonances, where \(K_c^{(-)}\) can be tuned continuously from \(+\infty\) to \(-\infty\).

\subsection{Scattering properties and bound states}

\begin{figure}[htp]
\centering
\includegraphics[width=12cm]{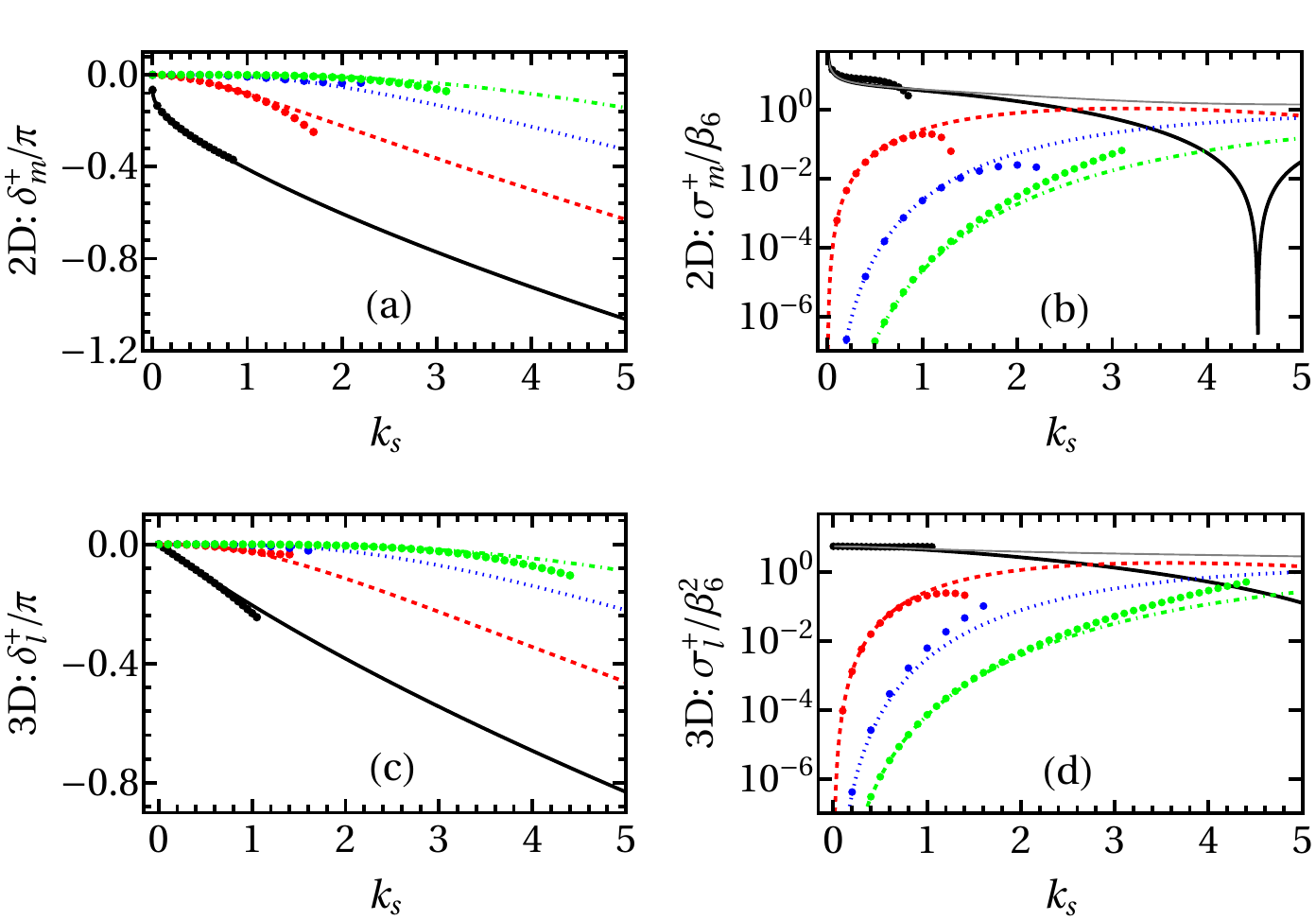}   
\caption{(Color online) Phase shifts and cross sections for the first four partial waves in 2D (a--b) and 3D (c--d) for purely repulsive potentials. The black solid, red dashed, blue dotted, and green dash-dotted lines correspond to \(s\)-, \(p\)-, \(d\)-, and \(f\)-waves, respectively. Thin gray lines indicate the total cross sections summed over these four partial waves. The dotted symbols represent results from the low-energy expansions of the scattering phase shifts [Eqs.~(\ref{low_E_ps_p_m0})--(\ref{low_E_ps_p_l3})]. The low-energy expansions of cross sections are omitted due to their lengthy expressions. The scaled momentum is defined as \(k_s = k \beta_6\).}
\label{Repulsive_scattering_plot}
\end{figure}
For the purely repulsive case, we analyze the scattering phase shifts by setting \(K_c^{(+)} = 0\) in Eqs.~(\ref{ps_de1})--(\ref{ps_de2}). This leads to the simplified expressions:
\begin{eqnarray}
\mbox{3D:} \quad \tan\delta_l^{+} &=& -\frac{Z_{12}^{(+)} }{Z_{11}^{(+)}}, \label{ps_de3} \\
\mbox{2D:} \quad \tan\delta_m^{+} &=& -\frac{Z_{11}^{(+)} + Z_{12}^{(+)} }{Z_{11}^{(+)} - Z_{12}^{(+)} }. \label{ps_de4}
\end{eqnarray}

In Fig.~\ref{Repulsive_scattering_plot}, we plot the phase shifts and cross sections for the first four partial waves in 2D and 3D. The phase shifts start from zero and increase with {\color{black}the scaled momentum \(k_s = k \beta_6\)}. Notably, each time the phase shift crosses an integer multiple of \(\pi\), the corresponding cross section exhibits a dip due to destructive interference (see, for instance, the black curve in panel (b)).

{\color{black}At low energies (\(k_s \ll 1\)), the scattering is dominated by the \(s\)-wave contribution, while higher partial waves become relevant only at elevated energies. In the 3D limit \(k \rightarrow 0\), the total scattering cross section asymptotically approaches a constant:
\[
\sigma_{\mathrm{tot}}^{\mathrm{3D}} = \sigma_{l=0}^{\mathrm{3D}} + \sigma_{l=1}^{\mathrm{3D}} + \sigma_{l=2}^{\mathrm{3D}} + \sigma_{l=3}^{\mathrm{3D}}+\sum_{j\geq 4}\sigma_{j}^{\mathrm{3D}} \approx \sigma_{l=0}^{\mathrm{3D}} \rightarrow \pi \left[ \frac{\Gamma(3/4)}{\Gamma(5/4)} \right]^2,
\]
where the low-energy expansions of the partial-wave contributions are given by
\begin{eqnarray}
    \sigma_{l=0}^{\mathrm{3D}} &= \frac{\pi \, \Gamma \left(3/4\right)^2}{\Gamma \left(5/4\right)^2}
    - \frac{ \pi \, \Gamma \left(3/4\right)^4}{12\, \Gamma \left(5/4\right)^4} \, k_s^2, \\
    \sigma_{l=1}^{\mathrm{3D}} &= \frac{\pi \, \Gamma \left(1/4\right)^2}{48 \, \Gamma \left(7/4\right)^2} \, k_s^4, \\
    \sigma_{l=2}^{\mathrm{3D}} &= \frac{4 \pi^3}{19845} \, k_s^6, \\
    \sigma_{l=3}^{\mathrm{3D}} &= \frac{4 \pi^3}{1715175} \, k_s^6.
\end{eqnarray}
In contrast, in the 2D case, the total cross section exhibits a logarithmic divergence as \(k \rightarrow 0\).} This divergence is a well-known feature of 2D scattering, arising due to the long-range logarithmic behavior of the 2D Green's function and the lack of a centrifugal barrier in the lowest channel.

We further compare exact numerical results with low-energy analytical expansions. The phase shifts in 2D are given by:
\begin{eqnarray}
\tan\delta_{m=0}^{+} &=& \frac{\pi}{2 \ln k_s + 3\gamma - 4\ln 2}, \label{low_E_ps_p_m0} \\
\tan\delta_{m=1}^{+} &=& \frac{\pi k_s^2 \left[ (19 - 36\gamma + 24 \ln(2/k_s)) k_s^2 - 48 \right]}{384}, \label{low_E_ps_p_m1} \\
\tan\delta_{m=2}^{+} &=& \frac{\pi k_s^4 \left(24 \ln k_s + 36\gamma - 11 - 48 \ln 2 \right)}{3072}, \label{low_E_ps_p_m2} \\
\tan\delta_{m=3}^{+} &=& -\frac{\pi k_s^4}{1280}, \label{low_E_ps_p_m3}
\end{eqnarray}
and in 3D:
\begin{eqnarray}
\tan\delta_{l=0}^{+} &=& -\left[ \frac{2 \Gamma(3/4)}{\Gamma(1/4)} + \frac{\pi k_s^3}{15} \right] k_s, \label{low_E_ps_p_l0} \\
\tan\delta_{l=1}^{+} &=& -\left[ \frac{\Gamma(1/4)}{24 \Gamma(7/4)} - \frac{\pi k_s}{35} \right] k_s^3, \label{low_E_ps_p_l1} \\
\tan\delta_{l=2}^{+} &=& -\frac{\pi}{315} k_s^4, \label{low_E_ps_p_l2} \\
\tan\delta_{l=3}^{+} &=& -\frac{\pi}{3465} k_s^4, \label{low_E_ps_p_l3}
\end{eqnarray}
where \(\gamma\) is Euler's constant. The analytical expressions (dotted symbols) show excellent agreement with the numerical phase shifts (solid lines) in the low-energy regime, as illustrated in Fig.~\ref{Repulsive_scattering_plot}.

\begin{figure}[tp]
\centering
\includegraphics[width=15.6cm]{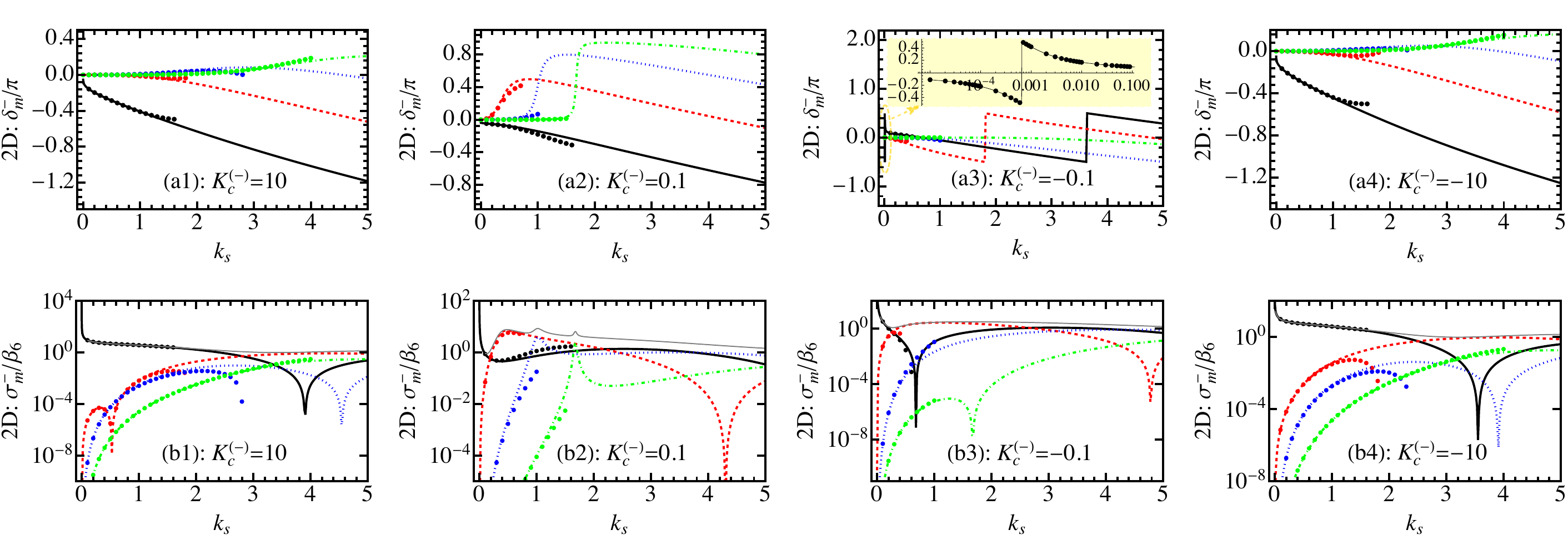}   
\caption{(Color online) Phase shifts and cross sections for the first four partial waves in 2D (a1--b4) and 3D (c1--d4), for potentials with attractive \textit{van der Waals} tails. The black solid, red dashed, blue dotted, and green dash-dotted lines correspond to \(s\)-, \(p\)-, \(d\)-, and \(f\)-waves, respectively. Thin gray lines show the total cross section summed over the first four partial waves. Dotted symbols indicate the results from low-energy expansions given in Eqs.~(\ref{low_E_ps_n_l0})--(\ref{low_E_ps_n_l3}). The explicit expressions for low-energy cross sections are omitted due to their complexity. The scaled momentum is defined as \(k_s = k \beta_6\).}
\label{Attractive_scattering_plot}
\end{figure}

\begin{figure}[t]
\centering
\includegraphics[width=11.5cm]{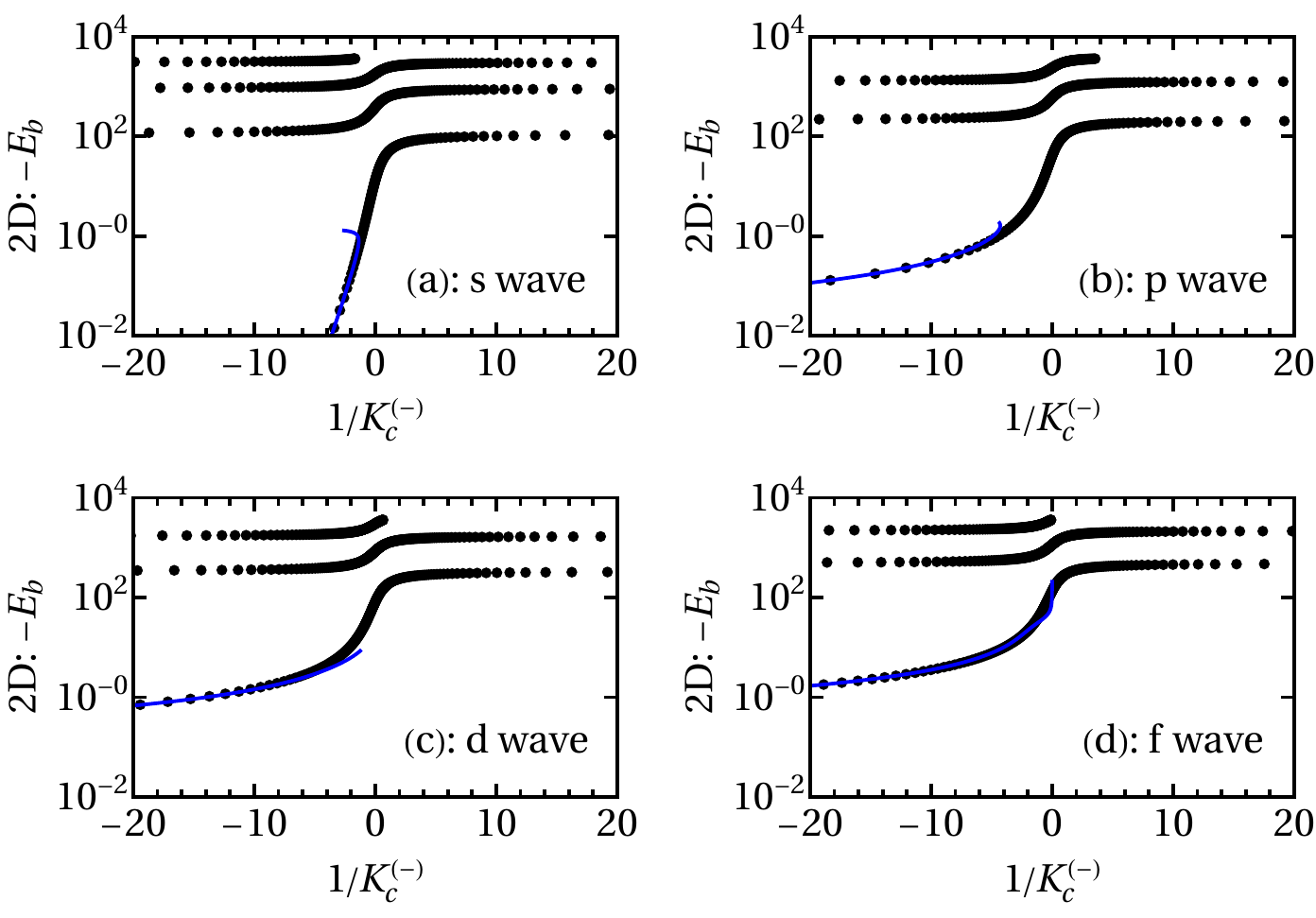}   
\caption{(Color online) Bound-state energies for the first four partial waves in 2D with attractive van der Waals tails. Black dotted symbols indicate exact numerical solutions, and blue solid lines represent low-energy expansions from Eqs.~(\ref{chi_m0})--(\ref{chi_m3}).}
\label{Attractive_bound_plot}
\end{figure}

For the attractive case in 2D, we investigate both scattering properties and bound-state spectra for the first four partial waves, as shown in Figs.~\ref{Attractive_scattering_plot} and~\ref{Attractive_bound_plot}. To illustrate the variation across resonance, we choose four representative values of the quantum defect parameter \(K_c^{(-)} = 10,\, 0.1,\, -0.1,\, -10\) in Fig.~\ref{Attractive_scattering_plot}.

In the \(s\)-wave channel, the cross section diverges as \(k \rightarrow 0\), consistent with the well-known logarithmic singularity in 2D. Moreover, a very sharp phase jump appears near a resonance in the case \(K_c^{(-)} = -0.1\), as shown in the inset of Fig.~\ref{Attractive_scattering_plot}(a3). Similar resonance-induced phase shifts are also observed in higher partial waves, such as in Fig.~\ref{Attractive_scattering_plot}(a2), corresponding to local maxima in the cross sections shown in panel (b2).

The low-energy expansions for the phase shifts in the attractive case are given by:
\begin{eqnarray}
\tan \delta_{m=0}^{-} &=& \frac{2\pi K_c^{(-)}}{A} - \frac{\pi^2 (K_c^{(-)2} + 1) k_s^2}{2 A^2}, \label{low_E_ps_n_l0} \\
\tan \delta_{m=1}^{-} &=& \frac{\pi k_s^2}{8 K_c^{(-)}}, \label{low_E_ps_n_l1} \\
\tan \delta_{m=2}^{-} &=& \frac{\pi (7 K_c^{(-)} - 4A) k_s^4}{2048 K_c^{(-)}}, \label{low_E_ps_n_l2} \\
\tan \delta_{m=3}^{-} &=& \frac{\pi k_s^4}{1280}, \label{low_E_ps_n_l3}
\end{eqnarray}
where \(A = 4 K_c^{(-)} \ln k_s + 6 \gamma K_c^{(-)} - 8 K_c^{(-)} \ln 2 - \pi\).

The energies of bound states are determined by Eq.~(\ref{chi_de}), where long-range properties are governed by the asymptotic behavior of the solutions, while short-range effects are encoded in the quantum defect parameter \(K_c^{(-)}\). In Fig.~\ref{Attractive_bound_plot}, we present the bound-state energies for the first four partial waves as functions of \(K_c^{(-)}\). The black dots represent numerical results, and the blue solid lines show analytical approximations based on the following low-energy expansions:
\begin{eqnarray}
\chi_{m=0}^{-} &=& \frac{\pi}{2B} + \frac{(4B^2 + \pi^2) k_s^2}{16 B^2}, \label{chi_m0} \\
\chi_{m=1}^{-} &=& \left(\frac{2B - 3 + 4\ln 2}{16}\right)\left(1 + \frac{\pi k_s^2}{8} \right)k_s^2 - \frac{\pi k_s^4}{768}, \label{chi_m1} \\
\chi_{m=2}^{-} &=& -\frac{k_s^2}{12} + \frac{\pi (7 - 8B) k_s^4}{4096}, \label{chi_m2} \\
\chi_{m=3}^{-} &=& -\frac{k_s^2}{32} + \frac{\pi k_s^4}{2560}, \label{chi_m3}
\end{eqnarray}
where \(B = 2 \ln {\color{black}|k_s|} + 3\gamma - 4 \ln 2\). From Fig.~\ref{Attractive_bound_plot}, we observe that the shallowest bound state emerges only in the \(s\)-wave channel as \(K_c^{(-)} \rightarrow {\color{black}0^{-}}\), while deeper bound states appear sequentially across all partial-wave branches as {\color{black}\(|K_c^{(-)}|\) increases}. All numerical results (dots) show excellent agreement with the low-energy analytical predictions (lines) in the weak-binding regime.

{\color{black}For attractive interactions in 2D, the QDT parameter \( K_c^{(-)} \) introduced in Eqs.~(\ref{low_E_ps_n_l0})-(\ref{low_E_ps_n_l3}) is directly related to the 2D scattering length \( a_{2D} \). Specifically, by comparing the low-energy behavior of the \( m = 0 \) phase shift,
\begin{equation}
\tan\delta_{m=0} = \frac{\pi}{2 \ln(k a_{2D})}, \qquad k \rightarrow 0.
\end{equation}
Comparing this behavior to that in Eq.~(\ref{low_E_ps_n_l0}), one can immediately obtain the relations
\begin{equation}
a_{2D}=e^{-\frac{\pi}{4K_c^{(-)}}}\bar{\beta}_6
\end{equation}
or
\begin{equation}
K_c^{(-)} = -\frac{\pi}{4 \ln(a_{2D}/\bar{\beta}_6)}
\end{equation}
with $\bar{\beta}_6 = e^{3\gamma/2}\beta_{6}/4$. Substituting this into Eq.~(\ref{low_E_ps_n_l0}), the \( s \)-wave phase shift can be expressed in terms of \( a_{2D} \) as
\begin{equation}
\tan\delta_{m=0}^{-} = \frac{\pi}{2 \ln(k a_{2D})} 
- \frac{\pi^2 + 16 \ln^2(a_{2D}/\bar{\beta}_6)}{32 \ln^2(k a_{2D})} \, k_s^2.
\end{equation}

Likewise, the low-energy bound state energy can also be expressed via \( a_{2D} \). In the limit \( a_{2D} \rightarrow \infty \), the \( s \)-wave binding energy reads
\begin{eqnarray}
E_b = -\frac{\hbar^2 \kappa^2}{m}, ~~~\kappa a_{2D} \simeq  1 -\left[\frac{1}{\pi}\ln^{2}\left(\frac{a_{2D}}{\bar{\beta}_{6}}\right)+\frac{\pi}{16}\right]\left(\frac{\beta_{6}}{a_{2D}}\right)^{2}.
\end{eqnarray}}

\section{Conclusions} \label{conclusion}
In this work, we have provided exact solutions to the 2D and 3D Schrödinger equations with isotropic van der Waals interactions of the form \(\pm 1/r^6\), using the generalized Neumann expansion method. We have also performed a detailed analysis of the asymptotic behavior of these solutions in both the short- and long-range limits. Based on these exact solutions, we constructed quantum defect theories for two-body systems with van der Waals-type potential tails in both quasi-2D and 3D geometries. Scattering properties and two-body bound states were thoroughly investigated through exact numerical calculations as well as low-energy expansion formulas. Our results not only provide a comprehensive understanding of two-body physics with long-range van der Waals interactions, but also offer useful input for future many-body studies involving \(\pm 1/r^6\) potentials. Furthermore, the analytical framework developed here can be extended to systems with other classes of $1/r^n$ long-range interactions. 
{\color{black}However, some widely used short-range potentials, such as the Lennard-Jones potential, do not permit the variable transformation that reduces the Schrödinger equation to the standard form of Eq.~(\ref{xfde}), which is essential for the analytical treatment applied in this work. Therefore, these cases lie beyond the scope of the current method and would require the development of new mathematical techniques or hybrid numerical-analytical approaches for their solution.}
\\

{\it Acknowledgements.} This work was supported by the National Key R\&D Program of China under Grant No.~2022YFA1405301 (R.Q.), the National Natural Science Foundation of China under Grant No.~12022405 (R.Q.) and No. 12104210 (J.J.), the Beijing Natural Science Foundation under Grant No.~Z180013 (R.Q.), the Natural Science Foundation of Top Talent of SZTU (GDRC202202, GDRC202312), and the Guangdong Provincial Quantum Science Strategic Initiative (No. GDZX2305006).

\appendix
{\color{black}
\section{Derivation of the uniform radial equation (Eq. (\ref{Schrodinger})) for both two-dimensional and three-dimensional systems}\label{ascasd}
We consider the long-range van der Waals-type interactions, which has a a central potential of the form
\begin{equation}
V(r) = \delta \frac{C_6}{r^6},
\end{equation}
where \( C_6 > 0 \) characterizes the strength of the interaction, and the index $\delta$ distinguishes between repulsive ($\delta = +1$) and attractive ($\delta = -1$) potentials. The full time-independent Schrödinger equation reads
\begin{equation}
\left[ -\frac{\hbar^2}{2\mu} \nabla^2 + V(r) \right] \Psi(\vec{r}) = \epsilon \Psi(\vec{r}),
\end{equation}
where \( \mu \) is the reduced mass and \( \Psi(\vec{r}) \) is the total wavefunction. Due to spherical symmetry of van der Waals-type interactions, we can employ the method of separation of variables to decouple the radial and angular degrees of freedom. Although the separation procedures differ fundamentally between 3D and  2D systems, the resulting radial equations can be cast into a unified form. In the following, we present a detailed analysis of both the 3D and  2D cases.

For 3D system, the Laplacian in spherical coordinates takes the form
\begin{equation}
\nabla^2 = \frac{1}{r^2} \frac{\partial}{\partial r} \left( r^2 \frac{\partial}{\partial r} \right)
- \frac{\hat{L}^2}{\hbar^2 r^2},
\end{equation}
where \( \hat{L}^2 \) is the square of the orbital angular momentum operator. This allows us to separate the total wavefunction as
\begin{equation}
\Psi(\vec{r}) = R_{\epsilon l}(r) Y_{ l m}(\theta, \phi),
\end{equation}
with \( Y_{ l m} \) denoting the spherical harmonics and \( R_{\epsilon l}(r) \) the radial wavefunction. Substituting this ansatz into the full equation and using the identity \( \hat{L}^2 Y_{ l m} = \hbar^2  l( l + 1) Y_{ l m} \), we obtain the radial equation:
\begin{equation}
\left[ -\frac{\hbar^2}{2\mu} \left( \frac{d^2}{dr^2} + \frac{2}{r} \frac{d}{dr} - \frac{ l( l+1)}{r^2} \right)
+ V(r) \right] R_{\epsilon l}(r) = \epsilon R_{\epsilon l}(r).
\end{equation}
To eliminate the first-order derivative term, we define the reduced radial function \( U_{\epsilon l}(r) = r R_{\epsilon l}(r) \), which leads to the simplified one-dimensional radial equation:
\begin{equation}
\left[ -\frac{\hbar^2}{2\mu} \frac{d^2}{dr^2} + \frac{\hbar^2  l( l+1)}{2\mu r^2} + V(r) \right] U_{\epsilon l}(r) = \epsilon U_{\epsilon l}(r).
\end{equation}
For analysis and numerical convenience, we introduce the scaled energy \( \bar{\epsilon} = 2\mu \epsilon/\hbar^2 \) and the van der Waals length $\beta_6 = (2\mu C_6 / \hbar^2)^{1/4}$, so that the radial equation for 3D case becomes dimensionless:
\begin{equation}\label{3DEq}
\left[ \frac{d^2}{dr^2} - \frac{ l( l+1)}{r^2} -\delta \frac{\beta_6^4}{r^6} +\bar{\epsilon} \right] U_{\bar{\epsilon} l}(r) = 0.
\end{equation}

For 2D systems with rotationally symmetric potentials \( V( r) \), the Laplacian in polar coordinate takes the form
\begin{equation}
    \nabla^2 = \frac{\partial^2}{\partial  r^2} + \frac{1}{ r} \frac{\partial}{\partial  r}
    + \frac{1}{ r^2} \frac{\partial^2}{\partial \varphi^2}.
\end{equation}
 where \(  r = \sqrt{x^2 + y^2} \) and \( \varphi \in (0, 2\pi) \) are the polar coordinates. Due to rotational invariance, the wavefunction can be separated as
\begin{equation}
\Psi( r, \varphi) = R_{\epsilon m}( r) e^{im\varphi},
\end{equation}
where m  is the angular momentum quantum number, and the operator \( \hat{L}_z = -i\hbar \partial/\partial \varphi \) commutes with the Hamiltonian. Substituting the ansatz into the full equation yields the radial equation
\begin{equation}
\left[ -\frac{\hbar^2}{2\mu} \left( \frac{d^2}{d  r^2} + \frac{1}{ r} \frac{d}{d  r} - \frac{m^2}{ r^2} \right)
+ V( r) \right] R_{\epsilon m}( r) =\epsilon R_{\epsilon m}( r).
\end{equation}
To eliminate the first-order derivative, we define the reduced radial wavefunction \( \chi_{\epsilon m}( r) = \sqrt{ r} R_{\epsilon m}( r) \), leading to a one-dimensional Schrödinger-like equation:
\begin{equation}
\left[ -\frac{\hbar^2}{2\mu} \frac{d^2}{d r^2}
+ \frac{\hbar^2 (m^2 - 1/4)}{2\mu  r^2} + V( r) \right] \chi_{\epsilon m}( r) = \epsilon \chi_{\epsilon m}( r).
\end{equation}
Apply the  the scaled energy \( \bar{\epsilon} = 2\mu \epsilon/\hbar^2 \) and the van der Waals length $\beta_6 = (2\mu C_6 / \hbar^2)^{1/4}$, we can obtain the dimensionless radial equation for 2D system:
\begin{equation}\label{2DEq}
    \left[ \frac{d^2}{dr^2} - \frac{ (m^2 - 1/4)}{r^2} -\delta \frac{\beta_6^4}{r^6} +\bar{\epsilon} \right] \chi_{\bar{\epsilon} m}(r) = 0.
\end{equation}

By comparing Eq.~(\ref{3DEq}) and Eq.~(\ref{2DEq}), we observe that the two radial equations share an identical mathematical structure. The only difference lies in the specific form of the centrifugal term, which can be unified via the mapping
\begin{equation}\label{lmrealtion}
 \left(m - \frac{1}{2}\right)\left(m + \frac{1}{2}\right) \longleftrightarrow  l( l + 1)
\quad \Rightarrow \quad  \quad m - \frac{1}{2} \rightarrow  l.
\label{AngularMapping}
\end{equation}
As a result, the analytical treatment of these two cases under the same potential proceeds in a formally analogous manner. Therefore, the 3D and  2D radial equations take the same dimensionless form (Eq. (\ref{Schrodinger})):
\begin{equation}\label{UnifiedRadialEq}
    \left[ \frac{d^2}{dr^2} - \frac{\bar{l}( \bar{l} + 1)}{r^2} -\delta \frac{\beta_6^4}{r^6} +\bar{\epsilon} \right] \bar{u}_{\bar{\epsilon} \bar{l}}(r) = 0.
\end{equation}
In 2D, $\overline{l} = m - 1/2$, where $m$ is the azimuthal quantum number, and $\bar{u}_{\bar{\epsilon} \bar{l}}(r) =\chi_{\bar{\epsilon} m}(r)$; in 3D, $\overline{l} = l$, where $l$ is the orbital angular momentum quantum number, and $\bar{u}_{\bar{\epsilon} \bar{l}}(r) =U_{\bar{\epsilon} l}(r)$. This unified formalism enables us to treat a wide class of central potentials in both two and three dimensions within the same analytical framework.
}

\section{Results of $W,Z$ coefficients}\label{WZ}

The asymptotic behaviors of the functions \(\overline{u}_{\overline{\epsilon}\overline{l}}^{1}\) and \(\overline{u}_{\overline{\epsilon}\overline{l}}^{2}\) as \(r \rightarrow +\infty\) are given by
\begin{eqnarray}
\overline{u}_{\overline{\epsilon}\overline{l}}^{1} &\rightarrow& \xi, \\
\overline{u}_{\overline{\epsilon}\overline{l}}^{2} &\rightarrow& \frac{\xi \cos(\pi \nu) - \eta}{\sin(\pi \nu)}, \label{u2asymlong}
\end{eqnarray}
where
\begin{eqnarray}
\xi &=& G(-\nu) r^{1/2} \lim_{r \to +\infty} J_{-2\nu}(\sqrt{\overline{\epsilon}} r), \label{xi} \\
\eta &=& G(\nu) r^{1/2} \lim_{r \to +\infty} J_{2\nu}(\sqrt{\overline{\epsilon}} r), \label{eta}
\end{eqnarray}
with the function
\begin{equation}
G(\nu) = \Delta^{-\nu} \frac{\Gamma(1 + \nu_0 + \nu)\Gamma(1 - \nu_0 + \nu)}{\Gamma(1 - \nu)} C(\nu),
\end{equation}
and \(C(\nu) = \lim_{j \to \infty} c_j(\nu)\).

The coefficients \(Z_{nj}^{\pm}\) [Eq.~(\ref{u1asymlongp})] and \(W_{nj}^{\pm}\) [Eq.~(\ref{u1asymlongn})] are dimensionless and universal functions of the scaled energy \(\Delta\), depending only on the angular momentum index \(\widetilde{l}\). Their explicit forms are given below.

\vspace{1em}
\noindent\textbf{Coefficients \(W_{nj}^{+}\):}
\begin{eqnarray}
W_{11}^{+} &=& \widetilde{X}^{-1} \left[ (i - \cot \pi \nu) G(-\nu)(-1)^{-\nu} + \frac{G(\nu)}{\sin \pi \nu}(-1)^{\nu} \right], \label{W11p} \\
W_{12}^{+} &=& 2 \widetilde{X}^{-1} (i - \cot \pi \nu) G(-\nu)(-1)^{-\nu} \sin(2\pi \nu), \label{W12p} \\
W_{21}^{+} &=& \widetilde{Y}^{-1} G(-\nu)(-1)^{-\nu}, \label{W21p} \\
W_{22}^{+} &=& 2 \widetilde{Y}^{-1} G(-\nu)(-1)^{-\nu} \sin(2\pi \nu). \label{W22p}
\end{eqnarray}

\vspace{1em}
\noindent\textbf{Coefficients \(W_{nj}^{-}\):}
\begin{eqnarray}
W_{11}^{-} &=& \frac{[\alpha \sin(\pi \nu) - \beta \cos(\pi \nu)] G(-\nu)(-1)^{-\nu} + \beta G(\nu)(-1)^{\nu}}{(X^2 + Y^2) \sin(\pi \nu)}, \label{W11n} \\
W_{12}^{-} &=& \frac{2[\alpha \sin(2\pi \nu) - \beta \cos(2\pi \nu) - \beta] G(-\nu)(-1)^{-\nu}}{X^2 + Y^2}, \label{W12n} \\
W_{21}^{-} &=& \frac{[\beta \sin(\pi \nu) + \alpha \cos(\pi \nu)] G(-\nu)(-1)^{-\nu} - \alpha G(\nu)(-1)^{\nu}}{(X^2 + Y^2) \sin(\pi \nu)}, \label{W21n} \\
W_{22}^{-} &=& \frac{2[\beta \sin(2\pi \nu) + \alpha \cos(2\pi \nu) + \alpha] G(-\nu)(-1)^{-\nu}}{X^2 + Y^2}. \label{W22n}
\end{eqnarray}

\vspace{1em}
\noindent\textbf{Coefficients \(Z_{nj}^{+}\):}
\begin{eqnarray}
Z_{11}^{+} &=& \widetilde{X}^{-1} \left[ \frac{G(\nu)}{\sin(\pi \nu)} \cos\left(\pi \nu - \frac{\widetilde{l} \pi}{2} - \frac{\pi}{4}\right) \right. \nonumber \\
&& \left. - (i - \cot \pi \nu) G(-\nu)(-1)^{\widetilde{l}} \sin\left(\pi \nu - \frac{\widetilde{l} \pi}{2} - \frac{\pi}{4}\right) \right], \label{Z11p} \\
Z_{12}^{+} &=& \widetilde{X}^{-1} \left[ \frac{G(\nu)}{\sin(\pi \nu)} \sin\left(\pi \nu - \frac{\widetilde{l} \pi}{2} - \frac{\pi}{4}\right) \right. \nonumber \\
&& \left. - (i - \cot \pi \nu) G(-\nu)(-1)^{\widetilde{l}} \cos\left(\pi \nu - \frac{\widetilde{l} \pi}{2} - \frac{\pi}{4}\right) \right], \label{Z12p} \\
Z_{21}^{+} &=& \widetilde{Y}^{-1} G(-\nu)(-1)^{\widetilde{l}+1} \sin\left(\pi \nu - \frac{\widetilde{l} \pi}{2} - \frac{\pi}{4}\right), \label{Z21p} \\
Z_{22}^{+} &=& \widetilde{Y}^{-1} G(-\nu)(-1)^{\widetilde{l}+1} \cos\left(\pi \nu - \frac{\widetilde{l} \pi}{2} - \frac{\pi}{4}\right). \label{Z22p}
\end{eqnarray}

\vspace{1em}
\noindent\textbf{Coefficients \(Z_{nj}^{-}\):}
{\setlength{\mathindent}{0pt}
\begin{eqnarray}
Z_{11}^{-} &=& \frac{-(-1)^{\widetilde{l}} [\alpha \sin(\pi \nu) - \beta \cos(\pi \nu)] G(-\nu) \sin\left(\pi \nu - \frac{\widetilde{l} \pi}{2} - \frac{\pi}{4}\right) + \beta G(\nu) \cos\left(\pi \nu - \frac{\widetilde{l} \pi}{2} - \frac{\pi}{4}\right)}{(X^2 + Y^2) \sin(\pi \nu)}, \label{Z11n} \nonumber \\
\\
Z_{12}^{-} &=& \frac{-(-1)^{\widetilde{l}} [\alpha \sin(\pi \nu) - \beta \cos(\pi \nu)] G(-\nu) \cos\left(\pi \nu - \frac{\widetilde{l} \pi}{2} - \frac{\pi}{4}\right) + \beta G(\nu) \sin\left(\pi \nu - \frac{\widetilde{l} \pi}{2} - \frac{\pi}{4}\right)}{(X^2 + Y^2) \sin(\pi \nu)}, \label{Z12n} \nonumber \\
\\
Z_{21}^{-} &=& \frac{-(-1)^{\widetilde{l}} [\beta \sin(\pi \nu) + \alpha \cos(\pi \nu)] G(-\nu) \sin\left(\pi \nu - \frac{\widetilde{l} \pi}{2} - \frac{\pi}{4}\right) - \alpha G(\nu) \cos\left(\pi \nu - \frac{\widetilde{l} \pi}{2} - \frac{\pi}{4}\right)}{(X^2 + Y^2) \sin(\pi \nu)}, \label{Z21n} \nonumber \\
\\
Z_{22}^{-} &=& \frac{-(-1)^{\widetilde{l}} [\beta \sin(\pi \nu) + \alpha \cos(\pi \nu)] G(-\nu) \cos\left(\pi \nu - \frac{\widetilde{l} \pi}{2} - \frac{\pi}{4}\right) - \alpha G(\nu) \sin\left(\pi \nu - \frac{\widetilde{l} \pi}{2} - \frac{\pi}{4}\right)}{(X^2 + Y^2) \sin(\pi \nu)}. \label{Z22n}\nonumber \\
\end{eqnarray}
}

{\color{black}
\section{Structure of $\nu$}\label{sec_nu}

\begin{figure}[htp]
\centering
\includegraphics[width=15.6cm]{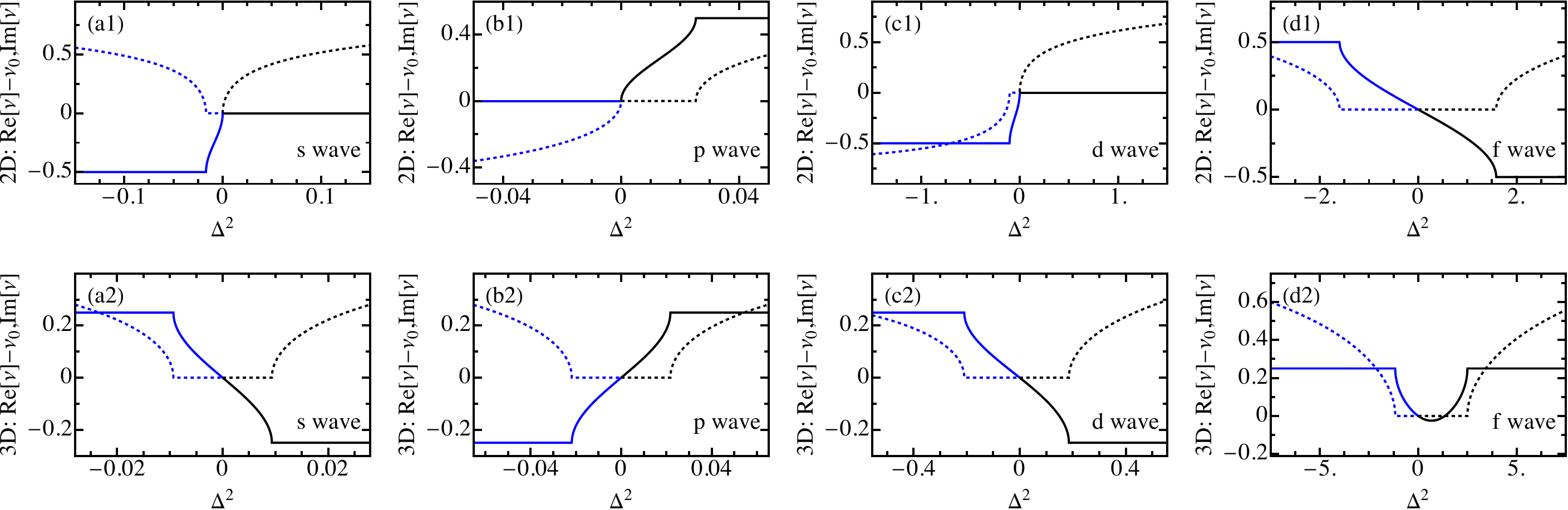}   
\caption{(Color online) Energy dependence of $\nu$ for the first four partial waves in 2D (a1--d1) and 3D (a2--d2). The black and blue curves correspond to the repulsive and attractive potentials, respectively. Solid lines denote the real part of $\nu$, while dashed lines indicate the imaginary part.}
\label{nuplot}
\end{figure}

As discussed above, a key step in constructing the exact solutions involves determining the value of $\nu$ for a given energy $\epsilon$ and angular momentum $\overline{l}$ by solving for the root of the characteristic function $\Lambda_{\overline{l}}(\nu, \varepsilon_s)$ in Eq.~(\ref{Lambdaeq}). On the complex plane, infinitely many branches of solutions for $\nu$ may exist. However, only one solution is physically relevant: the one that approaches $\nu_0$ as $\epsilon \rightarrow 0$ and varies continuously with increasing $|\epsilon|$~\cite{Gao1998C6,Jie_2016}. This root becomes complex beyond a critical scaled energy $\Delta_c$, which is listed in Table~\ref{table_nu} for the first four partial waves in both 2D and 3D.

In Fig.~\ref{nuplot}, we show the energy dependence of $\nu$ for the $s$-, $p$-, $d$-, and $f$-waves in 2D (a1--d1) and 3D (a2--d2). In 3D (a2--d2), each partial wave has a well-defined critical energy beyond which $\nu$ becomes complex, i.e., when $|\Delta^2| > |\Delta_c^2|$. In contrast, for 2D systems, $\nu$ can become purely imaginary even for the lowest partial waves. For general values of $l$ or $m$, the real part of $\nu$ tends to plateau as $|\Delta^2|$ increases beyond the corresponding critical energy. {\color{black}This plateau behavior implies that the structure of the wavefunction becomes simpler and more stable in this regime. The dominant behavior is governed by the long-range part of the potential, while the energy dependence of short-range contributions gradually diminishes.}

\begin{table}[t]
\centering
\caption{Critical scaled energy $\Delta_c^2 = (\overline{\epsilon}_c L^2 / 8)^2$ for the first four partial waves. Negative values of $\Delta^2$ correspond to repulsive potentials, while positive values correspond to attractive potentials.}
\label{table_nu}
\begin{tabular}{|c|c|c|c|c|}
\hline
              & \multicolumn{2}{c|}{2D}           & \multicolumn{2}{c|}{3D}           \\ \cline{2-5}
Partial Wave  & Repulsive         & Attractive     & Repulsive         & Attractive     \\ \hline
$s$           & $-0.0169$         & $0$            & $-0.00933$        & $0.00932$      \\ \hline
$p$           & $0$               & $0.0253$       & $-0.0218$         & $0.0217$       \\ \hline
$d$           & $-0.103$          & $0$            & $-0.209$          & $0.185$        \\ \hline
$f$           & $-1.60$           & $1.59$         & $-1.17$           & $2.50$         \\ \hline
\end{tabular}
\end{table}

These general behaviors can be obtained numerically. Nevertheless, additional analytical insights can be extracted in the low-energy limit by performing a small-$\Delta$ expansion of $\Lambda_{\overline{l}}(\nu, \Delta^2)$. Specifically, we obtain the following expansion for the root of the characteristic function in Eq.~(\ref{Lambdaeq}):
{\setlength{\mathindent}{20pt}
\begin{eqnarray}
\nu = \nu_0 - \frac{3 \Delta^2}{\nu_0 (1 - \nu_0^2)(1 - 4\nu_0^2)} - \frac{9 \left[8 - 7 \nu_0^2 (21 - 45 \nu_0^2 + 20 \nu_0^4)\right] \Delta^4}{4 \nu_0^3 (4 - \nu_0^2)(1 - \nu_0^2)^3 (1 - 4\nu_0^2)^3} + \mathcal{O}(\Delta^5),
\end{eqnarray} due to the mapping betweene expressed as:
\begin{eqnarray}
\nu_{m=0} &=& \frac{i}{16} \sqrt{\frac{3}{2}} \left(32 - 315 \Delta^2\right)\Delta + \mathcal{O}(\Delta^5), \\
\nu_{m=1} &=& \frac{1}{2} + \left(2 + \frac{119 \Delta^2}{9}\right)\Delta + \mathcal{O}(\Delta^5), \\
\nu_{m=2} &=& 1 + \left(i - \frac{\Delta}{48}\right)\Delta + \mathcal{O}(\Delta^3), \\
\nu_{m=4} &=& 2 + \left[\frac{3 - i}{120} + \left(\frac{377}{432000} + \frac{71i}{27000}\right)\Delta^2\right] \Delta^2+ \mathcal{O}(\Delta^5).
\end{eqnarray}
}
}

\end{document}